# Extreme ultraviolet spectrometer based on a transmission electron microscopy grid

Emily Sistrunk[1,2] and Markus Gühr[1,+]

[1]PULSE Institute, SLAC National Accelerator Laboratory, 2575 Sand Hill Road, Menlo Park, CA 94025, USA

[2]NIF and Photon Sciences, Lawrence Livermore National Laboratory, 7000 East Avenue, Livermore, CA 94550, USA

[+]*email: mguehr@stanford.edu*

We performed extreme ultraviolet spectroscopy using an 80 lines/mm TEM mesh as dispersive element. We present the usefulness of this instrument for dispersing a high harmonic spectrum from the 13[th] to the 29[th] harmonic of a Ti:Sapph laser, corresponding to a wavelength range from 60 to 27 nm. The resolution of the instrument is limited by the image size of the high harmonic generation region on the detector. The best resolution in first order diffraction is under 2 nm over the entire spectral range.

## I. Introduction

Extreme ultraviolet (EUV) light based on high harmonic generation[1–3] is finding an increasing array of applications from attosecond spectroscopy[4,5] and lensless imaging[6–9] to ultrafast element sensitive spectroscopy at the M-edges of 3d transition metals[10–12]. The radiation created by strong field high harmonic generations (HHG) provides a broadband spectrum in the vacuum ultraviolet (sometimes called extreme ultraviolet) which is discretized in the form of odd harmonics of the fundamental laser frequency. The optimization of the HHG process requires a quick estimate of the spectral content. However measuring the spectrum is much more complicated than in the optical domain. First, the radiation can only propagate in vacuum since air would otherwise absorb the light within a short distance. Second, in the EUV, reflection gratings only show high reflectivity under grazing incidence geometry. Third, two-dimensional area detectors need to be mounted in vacuum and their surface needs to be transparent for EUV photons. This requires a special back thinned CCD or multi-channel plates.

The work of Kornilov *et al.*[13] addresses these issues by showing a compact and efficient design for characterizing EUV light from a harmonic source. The spectral characterization is based on a special transmissive amplitude grating instead of a reflective grating. This lithographically produced element[14,15] exhibits line densities in the 10,000/mm range allowing for high dispersion and a compact spectrometer design. While this custom made grating is obviously a great tool for spectroscopy it is also rather expensive and fragile.

Here we propose the use of a commercial transmission electron microscope (TEM) grid as a grating for EUV light from high harmonic sources. These devices are mass produced, feature densities up to 660 per mm are commercially available for a very low price. We demonstrate sufficient resolution to distinguish different harmonics in the EUV spectrum down to a wavelength of 27 nm (photon energy of 45 eV) using a TEM grid of 80 lines/mm. The TEM grid can be inserted in many common beamline geometries requiring minimal modifications.

## II. Design

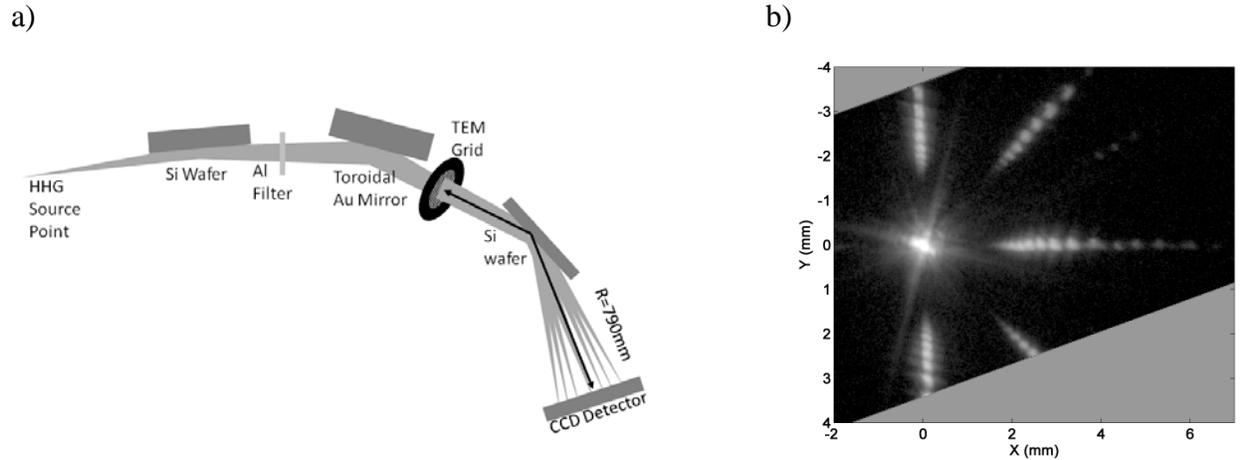

**Fig. 1: a)** Sketch of the setup. The EUV light is generated at the HHG source point, the Si wafer and Al filter suppress the generating laser but transmit the EUV light from 20-70 eV photon energy. The source point is imaged on the detector by a toroidal mirror. The TEM mesh with 80 l/mm is placed 790 mm in front of the CCD detector. The optical path contains a flat Si wafer in an interaction region for ultrafast experiments. **b)** The detected image (shown on a logarithmic grayscale) contains the bright zero order spot in the middle and half of the typical mesh diffraction pattern taken with polychromatic light of a high harmonic spectrum. The CCD detector was rotated with respect to the diffraction lines, which is corrected in the image.

Figure 1a shows our beamline layout. The EUV light is generated by strong field harmonic generation from a 30 fs, several mJ Ti:Sapph laser centered around a wavelength 800 nm and operating at a repetition rate of 1 kHz. For symmetry reasons, we only generate odd harmonics of the fundamental laser frequency, the wavelength such is 800 nm/k, with k being an odd number. The details of the realization of the harmonic generation process can be found in Ref.[16]. The infrared fundamental and EUV radiation are co-propagating and the much stronger infrared component is filtered using reflection off a silicon wafer[17] and a thin aluminum filter passage. The 100 nm thick aluminum filter is transparent in the photon energy range from 20 to 70 eV (62 to 18 nm), which means it passes harmonics 13 to 45 of the 800 nm fundamental. As shown recently, a multichannelplate can be used alternatively to filter the infrared light from the EUV[18]. The generation region of the EUV light is imaged without magnification on a detector by a grazing incidence toroidal mirror in 4f (in our case 4f=2.4m) geometry. A Si folding mirror (Si wafer) at the place of the interaction region for an EUV transient grating setup[19] is hit under grazing angle 22 deg,. The detector is a back-thinned, EUV sensitive CCD camera with (27.6 x 6.9) mm$^2$ active area and a pixel size of (26x26) μm$^2$.

The TEM grid is inserted at $R$=790 mm in front of the detector and attached to a gate valve with mounting hole for optical windows or filters. The grid is a model GS2000HS purchased from SPI supplies and is fabricated from copper. The mesh structure has a bar width of 5 μm, a hole width of w=7.5 μm and a pitch of 12.5 μm. The pitch, corresponding to 80 lines/mm was independently checked using the diffraction image of a He-Ne laser.

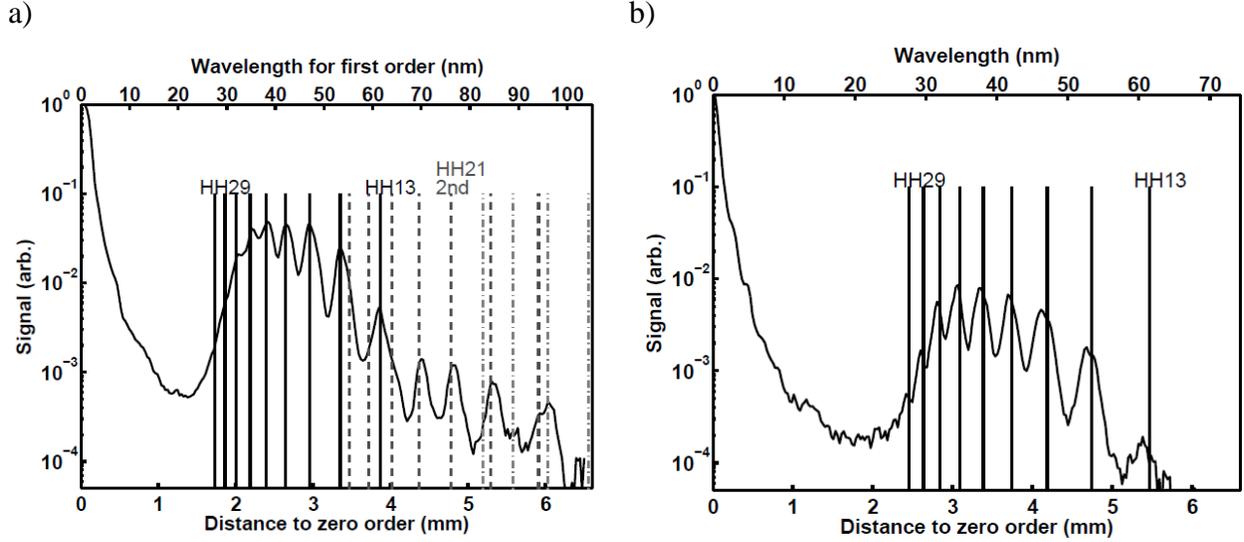

**Fig. 2: a)** Cut along the x-axis in Fig. 1b (black spectrum). We show the expected value for the diffraction peaks of the different high harmonics (from 13 to 29) in first order using a mesh pitch of 12.5 µm as black lines. The second and third order diffraction expectations are shown as dashed gray and dot-dashed light gray lines, respectively. The wavelength scale on top of the graph refers to the first order diffraction. **b)** Cut along an angle of 45 deg. between x and y axis in Fig. 1b (black spectrum). Once more we show the expectation of the mesh diffraction for different harmonics.

## III. Results and Discussion

Figure 1b shows the CCD detector image with the EUV light hitting the TEM grid on a logarithmic grayscale. At the point (0,0) mm the image shows the bright non-diffracted light, which we refer to as zero-order light. The diffraction of the EUV light is clearly visible in terms of 4 stripes, two of them, respectively, on the x and y axes, the other two in an angle of 45 deg to the axes. Due to constraints in the vacuum tubes, we only could transport half of the diffraction features to the detector. The complete eightfold structure, expected from the diffraction of a coherent source on a mesh can be reconstructed by inversion at the (0,0) point. We first analyze the data of the stripes parallel to the x and y axis. Even in the raw image one clearly discerns different spots along these lines. We will show that they correspond to different high order harmonics of the 800 nm drive laser.

Figure 2a shows an integration along the abscissa from point (0,0) to point (6.6,0) with an integration width along the ordinate of ¼ mm. We chose a logarithmic representation of the signal as in Fig. 1b in order to show the details of higher order diffraction. We analyze the position of the different peaks in terms of the dispersion relation $sin\theta=n\lambda/d$, where the diffraction angle is $\theta$, $d$ is the pitch of 12.5 µm and n is the diffraction order (not to be confused with the harmonic order). The distance of the diffracted light from the zero order on the detector is called $D$ and it is given as $D=tan\theta\, R$, where R=790 mm is the distance from TEM grid to detector. The calculated positions of harmonics 13 to 29 in the first order are shown as black lines, fitting the peaks closest to the zero order very well. The first order diffraction defines the wavelength scale on the upper figure axis. The second order (n=2) diffraction of harmonics 17 to 29 are shown as grey dashed lines. The prediction of second order diffraction of harmonics 17 to 23 clearly fit peaks with much lower intensity compared to the first order diffraction. Higher harmonics in second order are visible as shoulders on first order peaks. We even discern some third order diffraction of the grating. The calculated positions for harmonic 23-29 are given as dash-dotted light gray lines and fit some smaller structures in the spectrum well corresponding to third order diffraction of harmonics 25 and 27.

Figure 2b shows the integration from point (0,0) mm in Fig. 1b in 45 deg direction towards point (4,4) mm. The width of the integration area is again ¼ mm. The diffraction image of a mesh with equal periodicity in both directions has equal periodicity along x and y axis of the detector[20]. Thus in the direction 45 deg. to the axis, the distance to the zero order of a diffraction spot increases by a factor of $\sqrt{2}$. The expected distance for first order diffractive peaks from Fig. 2a scaled by $\sqrt{2}$ is shown as black lines and fits the structure very well. Once more, the first order diffraction defines the wavelength scale in the upper figure axis.

The highest harmonics in first order diffraction are certainly better resolved in the diagonal integration shown in Fig. 2b compared to the integration along the x-axis shown the Fig. 2a. In addition, the second order diffraction in Fig. 2a shows better resolution compared to first order for the group of harmonics than can be clearly attributed. The observation hints at the fact that the resolution is limited by the spot size of the EUV on the detector, which is 150 µm full width half maximum. For higher dispersion (as along the diagonal or for second order) the resolution increases consequently. Harmonic 19 has a full width half maximum of 2.5 nm in Fig. 2a, in Fig. 2b the width is decreased to around 1.7 nm. For an infinitely small spot size on the detector, the resolution will be determined by the number of illuminated structures on the mesh. Our experimental results in the diagonal direction are a factor of 3 wider than this limit.

We now discuss the intensities observed in the diffraction pattern. The overall transmission of the grid is 36% resulting from the large bar width. Due to the convolution theorem, the diffraction intensity is given by the intensity of the Fourier transform of a single square hole multiplied by the Fourier transform of the grid periodicity[20]. The intensity of the Fourier transform of the single square opening in two dimensions as a function of the detector coordinates (X,Y) is given as

$$I(X, Y) = I_0 \left( \frac{\sin\left(\frac{\pi w X}{\lambda R}\right)}{\frac{\pi w X}{\lambda R}} \right)^2 \left( \frac{\sin\left(\frac{\pi w Y}{\lambda R}\right)}{\frac{\pi w Y}{\lambda R}} \right)^2$$

Here, w is the opening width (7.5 µm), $\lambda$ is the wavelength and R the distance from the grid to the detector. For any particular wavelength, the diffraction intensity of the first order compared to the zero order is predicted to be 25% in the X direction and 6.25 % in the diagonal direction. Both the Fourier transform of a single square opening as well as the diffraction from the pitch scale identically with λ. Thus, the diffraction is independent of the wavelength. We now compare the experimentally determined intensities to this relation. Since our source has an enormous bandwidth, we need to estimate the composition of the zero order radiation. The five strongest harmonics exceeding their neighbors by at least a factor of 3 are harmonics 15 to 23. We thus assume that the zero order radiation is composed of 5 equal parts of these harmonics. The strength of the zero order, which is normalized to 1 in Fig. 2 is thus 1/5=0.2. In the x-direction, the individual harmonics 15-23 have a signal of 0.05, which is 25% of the spectrally pure zero order, as predicted. In the diagonal direction, the signal strength is about 0.008 which corresponds to 4%. This is lower than the expected 6.25%, but in the light of the coarse estimate still reasonable.

## IV.    Summary

We show that TEM grids can be used to disperse a EUV spectrum produced by high harmonic generation of an intense infrared laser. The dispersion and diffracted intensity are in agreement with the expected diffraction of the TEM grid. The setup uses a toroidal mirror to image the high harmonic generation region on a detector. For our setup, the resolution obtained by the 80 lines/mm grating is determined by the EUV spot size on the detector. Due to its extremely low investment and flexibility we anticipate that these grids are a superb tool for quickly estimating the spectral shape of the high harmonic spectrum. In addition, the diffraction efficiency of the transmission grating is independent of the wavelength making calibration of the grating response obsolete. Obviously, the relatively low line density limits the dispersion and therefore the resolution to shorter wavelengths. TEM grids with higher feature density around 660 lines/mm are commercially available. For our setup, they would allow resolution of harmonic orders around 65, corresponding to wavelength of 12 nm and photon energies of 100 eV.

**Acknowledgements**

We acknowledge discussions and experimental help with/from J. Grilj, T.J.A. Wolf and M. Koch. M.G. acknowledges funding via the Office of Science Early Career Research Program through the Office of Basic Energy Sciences, U.S. Department of Energy. This work was supported by the AMOS program within the Chemical Sciences, Geosciences, and Biosciences Division of the Office of Basic Energy Sciences, Office of Science, U.S. Department of Energy.